# Plasmonic terahertz detection by a double-grating-gate field-effect transistor structure with an asymmetric unit cell

V.V. Popov[1,2,*], D.V. Fateev[1,2], T. Otsuji[3], Y.M. Meziani[4], D. Coquillat[5], W. Knap[5]

[1]*Kotelnikov Institute of Radio Engineering and Electronics (Saratov Branch), 410019 Saratov, Russia*

[2]*Saratov State University, 410012 Saratov, Russia*

[3]*Research Institute of Electrical Communication, Tohoku University, 2-1-1 Katahira, Aoba-Ku, Sendai 980-8577, Japan*

[4]*Dpto. de Fisica Aplicada, Universidad de Salamanca, Pza de la Merced s/n, 37008 Salamanca, Spain*

[5]*Laboratoire Charles Coulomb UMR5221, CNRS&Université Montpellier 2, 34095 Montpellier, France*

**Abstract**

Plasmonic terahertz detection by a double-grating gate field-effect transistor structure with an asymmetric unit cell is studied theoretically. Detection responsivity exceeding 8 kV/W at room temperature in the photovoltaic response mode is predicted for strong asymmetry of the structure unit cell. This value of the responsivity is an order of magnitude greater than reported previously for the other types of uncooled plasmonic terahertz detectors. Such enormous responsivity can be obtained without using any supplementary antenna elements because the double-grating gate acts as an aerial matched antenna that effectively couples the incoming terahertz radiation to plasma oscillations in the structure channel.

*) Electronic mail: popov_slava@yahoo.co.uk

Detection of terahertz (THz) radiation by plasmonic nonlinearities in a two-dimensional (2D) electron channel of the field-effect transistor (FET) was originally proposed in [1]. Resonant (frequency selective) [2,3] as well as non-resonant (broadband) [4,5] plasmonic detectors have been studied. The frequencies of the plasmon resonances in the FET channel with identical boundary conditions at different ends of the gate are given by [6]

$$\omega_p = n \frac{\pi}{w_{\text{eff}}} \sqrt{\frac{e^2 N^{(0)} d}{m^* \varepsilon \varepsilon_0}}, \qquad (1)$$

where $N^{(0)}$ is the equilibrium electron density in the channel, $d$ and $\varepsilon$ are the thickness and dielectric constant of the barrier layer, $\varepsilon_0$ is the electric constant, $w_{\text{eff}}$ is the effective length of the gated 2D electron channel (which is greater than a geometrical length of the gate, $w$, due to the electric-field fringing effect), $n$ is an integer, and $e$ and $m^*$ are the electron charge and effective mass, respectively. For symmetry reasons, only the plasmon modes with odd indexes $n$ can be excited by THz wave incident at normal direction to the FET-channel plane, whereas the plasmon modes with even indexes $n$ remain dark. Resonant detection takes place at THz frequencies $\omega = \omega_p$ for high quality factors of the plasmon resonance such as $\omega_p \tau \gg 1$, where $\tau$ is the electron scattering time. The non-resonant plasmonic detection takes place when the equilibrium electron density in the FET channel decreases at negative gate voltages $U_g < 0$ so that the inequality $\omega_p \tau \ll 1$ becomes valid for $U_g \to U_{\text{th}}$, where $U_{\text{th}}$ is the channel pinch-off threshold voltage.

A metal grating gate is an efficient broadband coupler between plasmons in the FET channel and THz radiation [7, 8]. The grating-gate FET detectors that have been demonstrated recently [9-10] mostly employ a photoconductive THz plasmonic response [11, 12], which needs applying DC drain bias current in the device channel. However, strong drain current causes large voltage drop across a long electron channel and, hence, different unit cells of a large-area grating-gate FET structure turn out to be under different effective gate voltages. As a result, the net responsivity of the grating-gate FET plasmonic detectors decreases for strong drain currents and remains quite moderate (below 300 mV/W [10]). For symmetry reasons, the photovoltaic THz response (with no DC drain bias current) can appear only if some asymmetry is introduced into the unit cell of the periodic grating-gate FET structure. [Weak photovoltaic THz response of the grating-gate FET structures with a symmetric unit cell reported in Ref. 9 and 13 appeared most probably due to some small uncontrollable asymmetry of those structures or/and due to asymmetric irradiation of the structures by THz beam.]

In this paper, we propose a concept of the plasmonic THz detector based on a double-grating-gate (DGG) FET structure with an asymmetric unit cell (A-DGG-FET structure) demonstrating strong photovoltaic response [14]. This concept is substantiated by simulating the photovoltaic response of such THz detector. The device structure is shown schematically in Fig. 1. The DGG is formed by two one-periodic coplanar metal sub-gratings: sub-grating 1 and sub-grating 2 with widths of the sub-grating fingers $w_1$ and $w_2$, respectively. Sub-grating 1 is laterally shifted in respect to sub-grating 2 so that two successive slits between the DGG fingers within the A-DGG-FET period are of different width, $s_1$ and $s_2$. This introduces a required asymmetry in the unit cell of a periodic A-DGG-FET structure. Different gate voltages $U_1$ and $U_2$ can be applied inter-digitally to sub-gratings 1 and 2, respectively. We assume the characteristic parameters of the A-DGG-FET structure similar to those reported in [7] for InAlAs/InGaAs/InP DGG-FET structure with a symmetric unit cell: the gate-to-channel separation is $d = 65$ nm, electron density in the channel is $N^{(0)} = 2.5 \times 10^{12}$ cm$^{-2}$ for $U_{1,2} = 0$, the electron scattering time is $\tau = 2.9 \times 10^{-13}$ s (which corresponds to a room temperature electron mobility 12000 cm$^2$/(V·s) in InAlAs/InGaAs/InP heterostructure [7]). Profile of the equilibrium 2D electron density in the channel is calculated as a function of the gate voltages $U_1$ and $U_2$ in the self-consistent electrostatic approach (assuming zero electron temperature in the channel) [15]. For the structure with above mentioned parameters, the calculated channel pinch-off threshold voltages are $U_{1,2}^{(\text{th})} \approx -3.23$ V. For any $U_1$ and $U_2$, the equilibrium electron density in 2D electron channel, $N^{(0)}$, is a periodic function of the coordinate along the channel (the $x$-coordinate) $N^{(0)}(x) = N^{(0)}(x+L)$, where $L$ is the DGG period. We assume that a plane electromagnetic (EM) wave with angular frequency $\omega$ of the THz range and the polarization of the electric field across the DGG fingers is incident upon the A-DDG-FET structure at normal direction in respect to the structure plane.

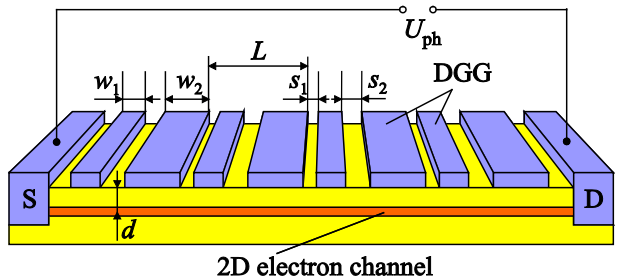

FIG. 1. (color online) Schematic view of the double-grating-gate FET structure with an asymmetric unit cell. External THz wave is incident normally from the top.

Plasmonic response in the FET channel originates from the nonlinear dynamics of 2D electron fluid described by the hydrodynamic equations [1]

$$\frac{\partial V(x,t)}{\partial t} + V(x,t)\frac{\partial V(x,t)}{\partial x} + \frac{V(x,t)}{\tau} + \frac{e}{m^*}E(x,t) = 0, \quad (2)$$

$$e\frac{\partial N(x,t)}{\partial t} - \frac{\partial j(x,t)}{\partial x} = 0, \quad (3)$$

where $E(x,t)$ is the in-plane electric field induced in the A-DGG-FET channel by the incident EM wave depending on the time $t$ and coordinate $x$ in 2D electron channel, $\tau$ is the electron scattering time, $j(x,t) = -eN(x,t)V(x,t)$ is the density of induced electric current, $N(x,t)$ and $V(x,t)$ are hydrodynamic electron density and velocity in 2D channel. There are two nonlinear terms in the system Eqs. (2) and (3): the second term in the Euler equation, Eq. (2), describes the nonlinear electron convection in 2D electron fluid and the product $N(x,t)V(x,t)$ defines the current density in the continuity equation, Eq. (3). Time average of the nonlinear current yields the detection signal [16].

Solution of the hydrodynamic equations, Eqs. (2) and (3), in the perturbation approach [11, 12] yields the rectified photocurrent density

$$j_{ph} = -e\sum_{q\neq 0} N_q^{(0)} V_{0,-q}^{(2)} - 2e\,\text{Re}\sum_{q\neq 0} N_{\omega,q}^{(1)} \left(V_{\omega,q}^{(1)}\right)^* \quad (4)$$

for zero DC drain bias current, where $N_{\omega,q}^{(1)}, V_{\omega,q}^{(1)}$ and $N_{0,q}^{(2)}, V_{0,q}^{(2)}$ are the amplitudes of the space-time Fourier harmonics of the induced electron density and velocity at the frequency $\omega$ of incoming THz radiation and at zero frequency, respectively, and $N_q^{(0)}$ are the amplitudes of the spatial Fourier harmonics of the equilibrium electron density $N^{(0)}(x)$ with $q = 2\pi p/L$ ($p = 0,1,2,3...$) being the reciprocal lattice vectors of the periodic A-DGG-FET structure. The corrections $N_{\omega,q}^{(1)}$ and $V_{\omega,q}^{(1)}$ are linear in the electric field amplitude of incoming THz wave, while the corrections $V_{\omega,q}^{(1)}$ and $V_{0,q}^{(2)}$ are proportional to the second power of the electric field amplitude of incoming THz wave in the perturbation approach. Hence, each sum-term in the right-hand side of Eq. (4) is proportional to the incoming THz power. For calculating the linear and nonlinear corrections to the electron density and electron velocity entering Eq. (4), one needs only the linear correction to the oscillating electric field induced in the structure channel by the incoming THz radiation, which can be calculated in a self-consistent linear electromagnetic approach described in [17]. The first-sum term arises in Eq. (4) only when the equilibrium electron density in 2D electron channel is spatially modulated along the $x$-direction. This term originates from the electron-convection nonlinear term in the Euler equation Eq. (2).

The photocurrent described by this term can be interpreted as a result of the plasma electrostriction effect in 2D electron channel with inhomogeneous electron density [12]. The second-sum term in Eq. (4) describes the plasmon-driven DC electron drag in 2D channel [11]. The both terms vanish in the DGG-FET structure with a symmetric unit cell.

In the open drain regime, the photocurrent, Eq. (4), generates the photovoltage between the drain and source contacts of the A-DGG-FET structure. This photovoltage, $U_{ph}$, can be calculated as $U_{ph} = -WRj_{ph}$, where $W$ is the width of the A-DGG-FET and $R$ is the device DC resistance depending on the gate voltages applied to the sub-gratings of the DGG (the A-DGG-FET structure is assumed to be 30 $\mu$m wide and 80 $\mu$m long). Then, the responsivity of the A-DGG-FET detector can be calculated by dividing the photovoltage by the THz power incident upon the device area.

Calculation show that the detection responsivity of the A-DGG-FET structure with a homogeneous 2D electron channel remains quite moderate (below 0.2 V/W) because there is no asymmetry in the channel of this structure while a remote asymmetric DGG produces only small THz photoresponse due to the plasmon induced electron drag in 2D electron channel. The asymmetry in the channel of the A-DGG-FET structure can be strongly enhanced by applying negative voltage to one of the two sub-gratings of the DGG. The detection responsivity grows linearly with increasing DC resistance of the A-DGG-FET channel and reaches 8 kV/W (for the asymmetry factors $s_1/s_2 < 0.5$) if the parts of 2D electron channel under the fingers of the biased sub-grating are strongly depleted (Fig. 2). We can not simulate the structure with even stronger depleted portions of the channel because the simulation procedure diverges for stronger depletion. However, since we do not see a saturation of the responsivity value with depletion of the channel in our modeling (notice that the saturation of the responsivity is inherent in a single-gate FET [18]), one can expect that the responsivity can be even higher for more negative gate voltages. Such enormous photoresponse of the A-DGG-FET THz detector is ensured by several critical factors. First of all, the A-DGG-FET detector combines the advantages of both the resonant and non-resonant plasmonic THz detectors. Strong THz photocurrent is generated by a nonlinear behavior of the plasmon mode resonantly excited with a quality factor $\omega_p \tau \gg 1$ in undepleted portions of 2D electron channel ($\omega_p \tau \approx 4$ for the first plasmon resonance at 2.15 THz and $\omega_p \tau \approx 8$ for the second plasmon resonance at 4.3 THz in Fig. 2). It is worth noting that the plasmon modes with both odd and even indexes $n$ are excited due to strong asymmetry of the A-DGG-FET structure (see black triangle symbols at the ordinate axis in Fig. 2). No plasmon modes can be excited in the depleted regions of the channel because of a low quality factor $\omega_p \tau \lesssim 1$. However, strong depleting of those regions greatly enhances the channel resistance, which leads to enormous enhancement of the photovoltage induced between the source and drain contacts of the entire A-DGG-FET structure. Note that the channel resistance per the unit cell of the A-DGG-FET structure for the depletion level corresponding to Fig. 2 is 5520 Ohm so that the value of the RC time constant remains quite short (less than 0.1 ns for typical inner capacitances of the unit cell 10-20 fF [6]). Secondly, the plasmon mode is injected into the plasmonic microcavity under the unbiased sub-grating finger predominantly from one end of the microcavity where the slit between this finger and an adjacent finger of the other sub-grating of the DGG is narrower. This is because stronger near electric field is induced by the incident THz wave at narrower slits [8]. Thirdly, the ungated parts of the channel are very important because they act as coupling elements synchronizing plasma oscillations under different fingers of the DGG. Finally, both the plasma electrostriction and electron drag mechanisms contribute strongly to the overall photoresponse in the A-DGG-FET due to its strong asymmetry. The calculations show that the responsivity grows steeply with increasing the electron mobility in the channel, reaching 35 kV/W and more for cryogenic temperatures.

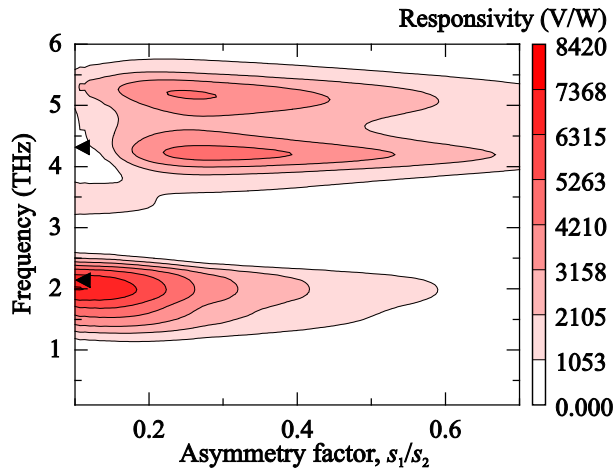

FIG. 2. (color online) Calculated responsivity of the A-DGG-FET detector as a function of THz frequency and the asymmetry factor, $s_1/s_2$, for $U_1 - U_1^{(th)} = 0.3$ mV for $U_2 = 0$. Parameters of the DGG are $L$ = 1300 nm, $w_1$ = 600 nm, $w_2$ = 500 nm. Black triangles at the ordinate axis mark the frequencies of the fundamental ($n$ = 1) and the second-order ($n$ = 2) plasmon modes calculated by Eq. (1) for $w_{\text{eff}}$ = $w_2$+100 nm.

At finite temperature, the electron density in 2D electron channel does not vanish for $U_{1,2} = U_{1,2}^{(th)}$ due to its thermal activation. In this case, the dependence of the equilibrium electron density in 2D electron channel on the gate voltage has more complex form [18]

$$N^{(0)}(U_g) = N^* \ln\left[1 + \exp\left(\frac{eU_0}{\eta k_B T}\right)\right], \quad (5)$$

where $N^* = C\eta k_B T/e^2$, $C$ is the gate capacitance per unit area, and $\eta$ is the so called ideality factor [19], $k_B$ is the Boltzmann constant, $U_0 = U_g - U_g^{(th)}$, and $T$ is the temperature. Accounting for finite temperature slightly changes the operating value of the gate voltage but does not reduce the high responsivity of the A-DGG-FET THz detector. While the electrostatic theory (in assumption of zero temperature) yields the equilibrium electron density $2.3 \times 10^8$ cm$^{-2}$ for $U_0 =$ 0.3 mV [which corresponds to Fig. 2], Eq. (5) gives the same equilibrium electron density for $U_0 = -0.27$ V for $T = 300$ K.

In conclusion, we have theoretically shown that the A-DGG-FET plasmonic THz detector can exhibit a greatly enhanced responsivity exceeding 8 kV/W at room temperature without using supplemetary antenna elements coupling the detector to incoming THz radiation. These results open up possibilities for drastic improvement of the performance of the plasmonic THz detectors.

We are grateful to M. Dyakonov and S.A. Nikitov for valuable comments. This work has been supported by the Russian Foundation for Basic Research (Grant Nos.10-02-93120 and 11-02-92101) and by the Russian Academy of Sciences Program "Fundamentals of Nanotechnology and Nanomaterials." Y.M.M. acknowledges the support from the Ministry of Science and Innovation of Spain (Projects PPT-120000-2009-4 and TEC-2008-02281) and the Ramon y Cajal program. The important help of the JST-ANR Japan-France International Strategic Collaborative Research Program "Wireless communication using Terahertz plasmonic-nano ICT devices" (WITH) is also aknowledged. The work was performed under umbrella of the GDR-I project "Semiconductor Sources and Detectors for Terahertz Frequencies."